\documentclass[aps,prl,reprint,showpacs,superscriptaddress]{revtex4-1}
\usepackage{epsfig,graphicx,amsmath,amssymb}
\usepackage{natbib}
\usepackage[nooneline]{subfigure}
\usepackage{graphicx}
\usepackage{amsmath}
\usepackage{color}

\setcitestyle{square}
\usepackage{subfigure}

\usepackage{enumitem}

\begin{document}

\title{Fluctuations of ring
polymers}
\author{Shlomi Medalion}
\affiliation{Department of Physics, Institute of Nanotechnology and Advanced
Materials, Bar-Ilan University, Ramat-Gan,52900, Israel}
\author{Erez Aghion}
\affiliation{Department of Physics,  Bar-Ilan University, Ramat-Gan,
52900, Israel}
\author{Hagai Meirovitch}
\affiliation{Department of Physics,  Bar-Ilan University, Ramat-Gan,
52900, Israel}
\author{Eli Barkai}
\affiliation{Department of Physics, Institute of Nanotechnology and Advanced
Materials, Bar-Ilan University, Ramat-Gan,52900, Israel}
\email{barkaie@biu.ac.il}
\author{David A. Kessler}
\affiliation{Department of Physics,  Bar-Ilan University, Ramat-Gan,
52900, Israel}

\pacs{05.40.Jc,02.50.-r,02.50.Ey,05.40.-a,05.10.Gg,05.20.Gg,36.20.Ey,36.20.Hb}
\date{\today}

\begin{abstract}
We present an exact solution for the distribution of sample averaged
monomer to monomer distance of ring polymers. 
For  non-interacting and   weakly-interacting models
these distributions correspond to the distribution of the area under
the reflected Bessel bridge and the Bessel excursion respectively, and
are shown to be identical in dimension $d \ge 2$.
A symmetry of the problem reveals that dimension $d$ and $4-d$ are
equivalent, thus the celebrated Airy distribution describing the areal
distribution of the $d=1$ Brownian excursion describes also a 
polymer in three
dimensions. 
For a
self-avoiding polymer in dimension $d$ we find numerically that the fluctuations
of the scaled averaged distance are nearly identical in dimension
$d=2,3$ and are well described to a first approximation by the non-interacting
excursion model in dimension $5$.
\end{abstract}


\maketitle

The statistical mechanics of polymers has been well studied for many years
due both to the numerous practical applications of polymers as well as their many interesting properties. 
One signal finding is that the overall size of a polymer of length $N$ scales as
$N^{\nu_p}$, where $\nu_p$ is a dimension dependent critical exponent. This is reflected in the behavior of various
observables, such as  the average end-to-end distance,
and the radius of gyration ($R_g$) 
\cite{flory1969statistical,de1979scaling}. The scaling exponent is known to be sensitive to  the excluded volume
interaction applied to part or all of the monomers. 
Other geometrical constraints applied to the chain,
such as cyclization (where the first monomer is connected to the last one), leading to
a ``ring'' polymer, affect only the prefactor for these quantities. Ring polymers are commonly found in many biological
systems e.g., bacterial and mitochondrial genomes, as well as DNA plasmids used
in many molecular biology experiments \cite{alberts2002molecular}. Recently,
ring polymers were also studied in the context of a model
for chromosome territories in the nucleus of eukaryotic cells
\cite{halverson2012rheology}.

The conformational fluctuations of some
polymer models can be analyzed using the theory of random walks (RW)
\cite{flory1969statistical,de1979scaling}. In particular, the fluctuations of 
the polymer size are of
 physical and biological importance. In the current paper we study the 
 distribution of sample-averaged monomer-to-monomer distance of ring polymers, both for ideal, noninteracting polymers
 as well as for polymers with excluded volume. In particular, we exploit recent mathematical
development on $d$ dimensional
 constrained Brownian motion (defined below) 
\cite{kessler2014distribution}
to
find an exact expression 
for the distribution of sample
averaged monomer to monomer distance for both ideal ring polymers and those with an additional excluded volume interaction applied at a single point.
This observable yields insight on the sample averaged fluctuations of
polymer sizes. 
An important ingredient of this calculation
 is the identification of the appropriate boundary conditions for 
the underlying equation, a variant of the Feynman-Kac
equation, which 
depends on  both the dimensionality and the interaction, 
and in turn yields
a selection rule for the  solution.  The resulting distributions are then compared numerically to those measured in simulations for a ring polymer with full excluded volume constraints.

As we have noted, constrained random walks lie at the heart of our analysis.  For rings, the primary constraint is that the path
returns to the origin after $N$ steps. 
 Statistics of such constrained one dimensional
 Brownian paths
have been the subject of much mathematical research~\cite{pitman1999brownian,janson2007brownian,perman1996distribution,pitman2001distribution}.
These constrained paths have been given various names, depending on the additional constraints imposed.
The basic case is that of a Brownian
bridge where the return to the origin is the sole constraint. For a Brownian excursion, 
the path is also forbidden from reaching
the origin in between. 
 Majumdar and Comtet used  Brownian excursions to determine statistical
 properties 
of the fluctuating 
Edwards Wilkinson interface in an interval 
 \cite{majumdar2004exact,majumdar2005airy}.
The focus of most previous work has been on the constrained one
dimensional Brownian paths which describes inherently non-interacting systems
(note that the problems of non-intersecting Brownian excursions
\cite{tracy2007nonintersecting}
or vicious random walkers \cite{schehr2008exact} in higher dimensions are
exceptions).  For the case of the fluctuations of ring polymers, we need to 
 extend the theory of Brownian excursions and bridges
 to other dimensions.  
We address the  influence of different kinds of
interactions on the polymer structure, both analytically (for a single point
interaction) and numerically (for a polymer with excluded volume interactions). 
These models yield rich physical behaviors and open new questions.\\

{\em Polymer Models.}
 We consider three lattice models of ring polymers with $N$ bonds, each of length $b$,  in $d$ dimensions.  
The simplest polymer model is an ``ideal
ring'' - a closed chain without excluded volume, 
where different monomers can occupy
the same lattice site. While such a polymer 
 does not exist in nature, its global
behavior is the same as that of a polymer at the Flory $\theta$-temperature
\cite{flory1969statistical}. An ideal ring chain corresponds exactly to an unbiased RW in $d$ dimensions which starts and ends at
the origin, i.e. a $d$-dimensional bridge. In the second model, the ``weakly interacting ring polymer", 
the first and last monomers are tied to the origin and no other monomer 
is allowed
to occupy this lattice site. This case is equivalent to that of $d$-dimensional excursions.
The third model is a ring polymer with excluded volume interactions, also
called a self-avoiding walk (SAW).  Further details on the polymer models and simulation methods 
are provided in the supplementary material (SM).
 We first consider the ideal and weakly interacting
ring models, for which we can provide an analytic solution.

{\em Bessel process.} 
In the analogy between statistics of  an ideal polymer and  a RW, the position 
of the $i$th monomer, ${\bf r}_i$,
corresponds to the position ${\bf r}_i$ of the random walker after $i$ time
steps  and $N$ is proportional to the total observation time. 
 The Bessel
process \cite{schehr2010extreme,martin2011first} describes the dynamics of the
distance $r = | {\bf
r} |$ from the origin of a Brownian particle in $d$ dimensions. 
This process is described by the following Langevin equation:
\begin{equation}
\dot{ r } = { ( d - 1) \over 2 r} + \eta(t),
\label{eq:Langevin}
\end{equation}
where $\eta(t)$ is  Gaussian white noise
satisfying $\langle \eta\rangle=0$ and $\langle \eta (t) \eta(t') \rangle = 
\delta(t - t')$.
One may map  the polymer models to
the Bessel process  using $\langle {\bf r}(t)^2 \rangle =
 d\,t = N b^2 = \langle {\bf R}^2 \rangle$, where ${\bf R}^2$ is the mean square
end-to-end distance of an ideal linear polymer chain without constraints.
In what follows we choose $b^2=d$ and  $t=N$.

{\em Bessel Excursions and Reflected Bridges.}
The process $r(t)$ with the additional constraint of starting and ending at the
origin, is called a
reflected (since $r\ge 0$) Bessel bridge. This process corresponds to an ideal
(non-interacting) ring chain. Bessel 
excursions are paths still described by Eq. (\ref{eq:Langevin}) however with the
additional constraint that any path that reaches the origin 
(besides  $t=0$
and $t=N$) is excluded.
The Bessel excursion corresponds to what we have called the ``weakly interacting" ring chain,
where a multiple occupation of the origin is not allowed.
The mapping of the polymer models to the Bessel process,
allows us to extract statistical properties of the former with new
tools developed in the stochastic community 
\cite{kessler2014distribution,majumdar2005airy,barkai2014area}.

{\em The Observable $A$.}
For a ring polymer, let ${\bf r}_i$ be the position of the $i$th monomer in
space where $i=0..N$, and we place the origin at the position of the zeroth monomer, $r_0$.
For the weakly interacting chain this monomer is also the excluding one.
We study  a new measure, $A$, for the size  of a ring polymer, defined by
\begin{equation}
A = \sum_{i=0} ^N | {\bf r}_{i} - {\bf r}_0| .
\label{eq:observableA}
\end{equation}
In the RW language, $A$ is the area under a random process, and hence is a random variable itself.
Clearly $\bar{l}= A/N$ is the sample averaged distance of the monomers from the
origin. 
Specifically, let the area under the random Bessel curve $r(t)$
be denoted by $A_B=\int_0 ^t r(t) {\rm d} t$ (the subscript $B$ is for
Bessel). 
More generally the mapping of the processes implies that in the limit
of large $N$ the distribution of $A_B/ \langle A_B \rangle$ is identical
to the distribution of $A / \langle A \rangle$ (or $\bar{l} / \langle
\bar{l} \rangle$), with the corresponding constraints.

\begin{figure}
\centering
\includegraphics[width=0.55\textwidth]{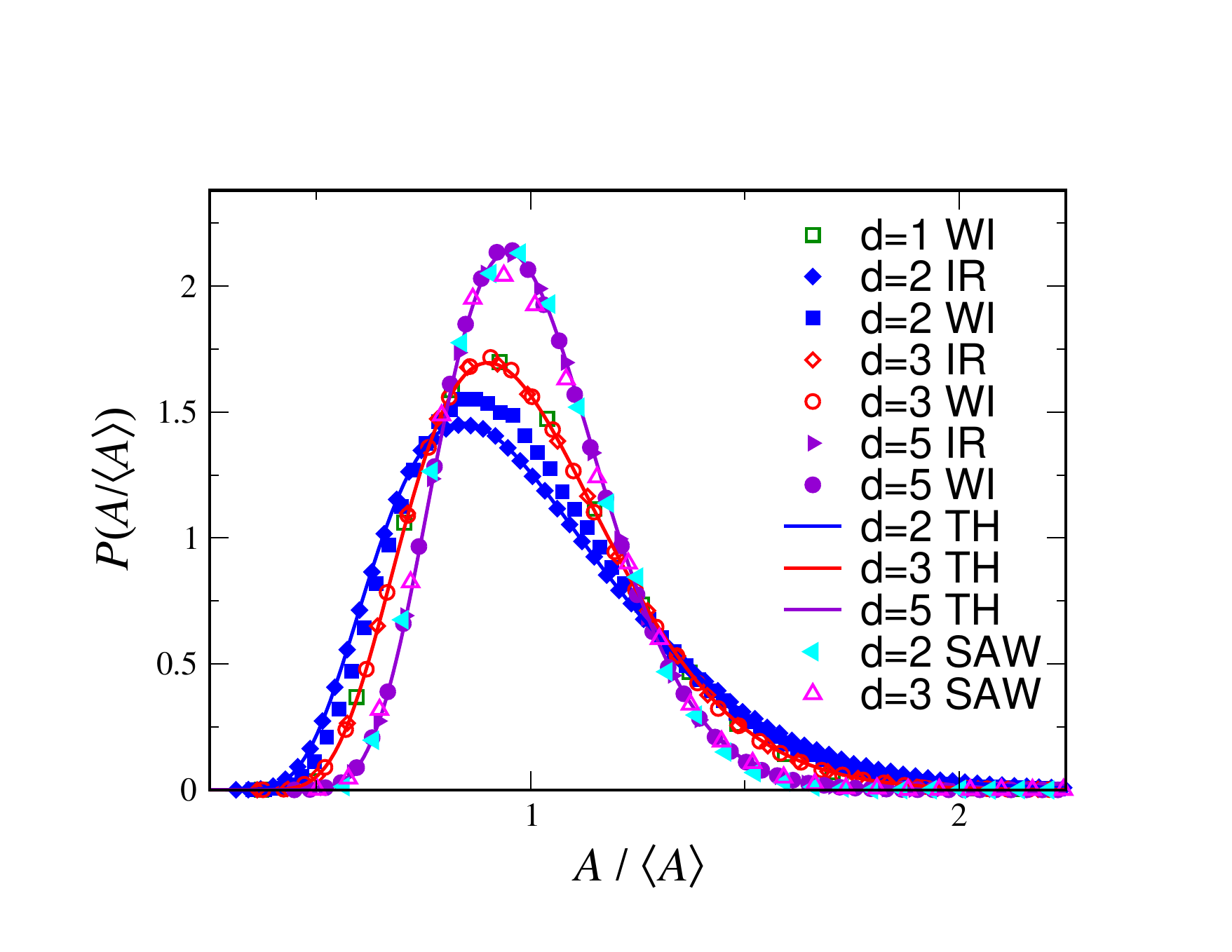}
\caption{(color online) Theoretical $P^{+}(A/\langle A\rangle)$ 
Eq.
(\ref{eqPDF})
in dimension $d=2,3,5$ nicely matches simulations of the ideal ring (IR) 
and the weakly interacting (WI) models, the exception being the weakly interacting
model in $d=2$ due to critical slowing down (the simulations did
not converge for $N=10^6$, see text). The theoretical curves (solid lines) for the two models are identical for $d \ge 2$. The $d=1$ WI theory
is identical to the $d=3$ theory, which the simulation confirms.
The SAW simulations in $d=2,3$ are practically
indistinguishable from each other and from the theoretical curve corresponding to 
a noninteracting ring in dimension $d=5$. 
}
\label{fig:allD_theory}
\end{figure}

{\em Numerical results.} In  
Fig. \ref{fig:allD_theory}
we plot the probability density function (PDF) of the
scaled  random
variable $A /\langle A \rangle$. Both results of simulations  and theory
are presented, however at this stage let us focus on the main features
as revealed in the simulations. 
For dimensions $d=2,\ldots,5$, there is a clear trend of narrowing of the PDF for increasing $d$. This trend is explained by examining Eq. (1): As $d$ increases, the noise term becomes negligible compared with the force term, resulting in smaller fluctuations and narrower tails.  Against this expected trend are the results in $d=1$ for the weakly interacting model. As we will show analytically, the weakly interacting model
in dimensions one and three surprisingly have the same distribution even though $d=1$ has a vanishing deterministic force term in Eq. (1) while for
d = 3 the force is clearly not zero.
In addition, we observe that for $d\ge 3$ the shape of the distribution of weakly interacting and ideal ring chains coincide, indicating
that weak interactions are negligible (when $N \to \infty$). As we shall see, this is also observed in the theory. Indeed the theory discussed below  suggests that these two distributions are already identical for $d=2$. However, since this is a critical dimension, due to extremely slow convergence, we don't see this behavior in the simulations. This asymptotic convergence  is logarithmic
(see SM) and an $\varepsilon$ expansion shows that it is reminiscent of critical slowing down.    As for the SAW polymer,
we see that the fluctuations are considerably reduced compared to the other models. This is due to the fact that the number of configurations of a SAW polymer is smaller
than for the other models, hence fluctuations are smaller. A striking 
observation is that the two and three dimensional SAW results are identical,
both being equal to the simulations of the $d=5$ models.
We now address these observations with theory.

{\em Functionals of Constrained Bessel Processes.}
Our goal is to find the PDF $P(A_B,t)$ of the
functional $A_B =\int_0 ^t r(t) {\rm d} t$  of the Bessel process, constrained
to start and end at the origin. 
We show that the difference between the weakly interacting model (the Bessel
excursion) and the ideal polymer
 (the reflected Bessel bridge) enters through the boundary condition 
in the Feynman-Kac type of equations describing these functionals.
The choice of boundary condition turns out to be non-trivial and controls
the solution. Other aspects of the solution 
follow the steps in \cite{kessler2014distribution}. 

It is useful to find first the Laplace transform of $P(A_B,t)$, i.e., 
$\tilde{P}(s,t)= \int_0 ^\infty P(A_B, t) \exp( - A_B s) {\rm d} A_B$ to solve
the equations, and invert $\tilde{P}(s,t)$ back to $P(A_B,t)$.
Let $G_t(r,A_B|r_0)$ be the joint PDF of the random pair $(r,A_B)$
with initial condition $G_0 (r, A_B| r_0) = \delta(A_B) \delta(r - r_0)$
and $\tilde{G} = \tilde{G}_t(r,s|r_0)$ its Laplace pair. The modified
Feynman-Kac
equation reads \cite{carmi2011fractional}:
\begin{equation}
{1 \over 2}  \left[ { \partial^2 \over \partial r^2} + {\partial \over \partial r} { 1 - d
\over r} \right] \tilde{G} - s r \tilde{G} 
= {\partial \over \partial t } \tilde{G}.
\label{eqFK}
\end{equation}
with $\tilde{G}|_{t=0} = \delta(r- r_0)$ and $r_0$ a cutoff which is eventually taken to zero. For $d=1$,  the second term on the
right hand side vanishes, and we get the celebrated Feynman-Kac equation 
corresponding to Brownian functionals \cite{brownianfuncMaj}. 
The third linear term $- s r \tilde{G}$ stems from the choice of
our observable, namely our functional $A_B$ is linear in $r$~\cite{carmi2011fractional}. 
Since we are describing a ring polymer,  the Bessel process must start and end
on the origin, and so, following \cite{majumdar2004exact}, we need to calculate
\begin{equation}
\tilde{P}(s,t) = \lim_{r=r_0 \rightarrow 0} {\tilde{G}_t (r, s|r_0) \over
\tilde{G}_t ( r, 0|r_0)}.
\label{eq:eqProbability}
\end{equation}
The denominator gives the proper normalization condition. 

The first step in the calculation is to perform a similarity transformation:
\begin{equation}
\tilde{G}_t (r, s | r_0) = \left( { r \over r_0} \right)^{ { d-1 \over 2}}
\widetilde{K}_t (r, s|r_0). 
\label{eq:Similarity}
\end{equation}
Using Eq. (\ref{eqFK}), $\tilde{K}_t ( r, s|r_0)$ is the imaginary time
propagator of a Schr\"odingier operator:
\begin{equation}
\hat{H} \widetilde{K}_t ( r , s| r_0) + {\partial \over \partial t} \tilde{K}_t ( r
, s | r_0 ) = \delta (r-r_0) \delta(t)
\label{eqSch}
\end{equation}
with the effective Hamiltonian:
\begin{equation}
\hat{H} = - {1 \over 2}  {\partial^2 \over \partial r^2} +  { (d -2)^2 - 1 \over 8 r^2} +
s r.
\label{eqH}
\end{equation}
The effective Hamiltonian reveals a subtle symmetry, namely two systems
in dimensions $d_1$ and $d_2$ satisfying $d_1+ d_2=4$ behave identically.
Note that this symmetry is not affected by the choice of functional (or
observable)
since the latter only modifies the last term in $\hat{H}$. This explains the identity of the
$d=1$ and $d=3$ PDFs noted earlier.

{\em Boundary Conditions for Ideal and Weakly Interacting Models.}
The solution of Eq. (\ref{eqSch})
\begin{equation}
\widetilde{K}_t(r,s|r_0) = \sum_k \phi_k(r) \phi_k(r_0) e^{-\lambda_k t}
\end{equation} 
 is constructed~\cite{kessler2014distribution} from the eigenfunctions $\phi_k$ of
 $\hat{H}$
where $\lambda_k$ is the $k$th eigenvalue and the normalization condition is
$ \int_0 ^\infty \phi_k ^2(r) {\rm d} r=1$. The subtle point in the
analysis is the
assignment of the appropriate boundary condition corresponding to the underlying polymer
models we consider. The eigenfunctions at small $r$ exhibit one of  two behaviors:
\begin{equation}
\phi^{+} _k \sim d^{+} _k r^{{1 + | 2-d|\over 2} } \  \ \mbox{or} \ \ \
\phi^{-} _k \sim
d^{-} _k r^{ {1 - |2 - d|\over 2}}.
\label{eqSmallr}
\end{equation}
From the normalization condition, the $\phi^{-}$ solution cannot be valid 
for $d\ge 4$ and $d \le 0$.
For the critical dimension  $d=2$ the two solutions are:
$\phi^{+} _k \sim d^{+} _k r^{{1 \over 2}} \  \ \mbox{or} \ \ \ \phi^{-} _k \sim
d^{-} _k r^{ {1 \over 2}} \ln r$.
%
We now solve the problem for the two boundary conditions and then
show how to choose the relevant one for the physical models under investigation. 

{\em  The distribution of $A/\langle A \rangle$}.
Following
the Feynman-Kac formalism described above and performing the inverse Laplace transform
~\cite{kessler2014distribution}, 
 we find two solutions for the PDF of the scaled variable $\chi \equiv A /\langle A\rangle$ 
\begin{widetext}
\begin{align}
 p^{\pm} (\chi) &=  -\frac{\Gamma(1\pm|\alpha|)}{2\pi \chi} 
\left(\frac{4}{(\sqrt{2} c_{\pm} \chi)^{2/3}}\right)^{\pm |\alpha|+1}
\nonumber\\
& \times \sum_{k=0}^\infty [\tilde{d}^{\pm} _k]^{2}
\bigg[\Gamma\left(\frac{5}{3}\pm|\nu|\right)\sin\left(\pi\frac{2\pm3|\nu|}{3}
\right){}_2F_2\left(\frac{8}{6}\pm\frac{|\nu|}{2},\frac{5}{6}\pm
\frac{|\nu|}{2};\frac{1}{3},\frac{2}{3};-\frac{2\lambda_k^3}{27
(c_{\pm} \chi)^2}\right)
\nonumber\\
& - \frac{\lambda_k}{(\sqrt{2} c_{\pm} \chi)^{2/3}}
\Gamma\left(\frac{7}{3}\pm |\nu|\right)\sin\left(\pi\frac{4\pm
3|\nu|}{3}\right){}_2F_2\left(\frac{7}{6}\pm\frac{
|\nu|}{2},\frac{5}{3}\pm\frac{
|\nu|}{2};\frac{2}{3},\frac{4}{3};-\frac{2\lambda
_k^3}{27 (c_{\pm} \chi)^2}\right) \nonumber\\
& + \frac{1}{2}\left(\frac{\lambda_k
}{
(\sqrt{2} c_{\pm} \chi)^{2/3}}\right)^2 \Gamma\left(3\pm |\nu|\right)\sin\left(\pm
\pi|\nu|\right){}_2F_2\left(2
\pm\frac{ |\nu|}{2},\frac{3}{2}\pm\frac{
|\nu|}{2};\frac{4}{3},\frac{5}{3};-\frac{2\lambda_k^3}{27
( c_{\pm} \chi)^2}\right)\bigg].
\label{eqPDF}
\end{align}
\end{widetext}
The solution is independent of $N$ and valid in the limit of $N \to \infty$. 
Here,  $|\alpha|= |d-2|/2$, $|\nu| = 2 |\alpha|/3$, and $_2F_2(\cdot)$ refers to the generalized hypergeometric functions.
The supplementary material
 provides a list of $\lambda_k$
and $d_k$ values for $d=1,\ldots 4$.
For $d=1$ the solution agrees with the known results\cite{majumdar2004exact,majumdar2005airy,shepp1982integral,knight2000moments}, where the $+$ solution is the celebrated Airy distribution~\cite{majumdar2004exact,majumdar2005airy}.
The average of $A$ is
\begin{equation}
\langle A \rangle^{\pm} = c_{\pm} N^{3/2}, \ \
c_{\pm}={  \pi \Gamma \left( \pm  {\left| 2 - d\right| \over 2}
 + { 3 \over 2} \right) \over 4\sqrt{2}  \Gamma\left( \pm { | 2 - d| \over 2} + 1
\right)
}. 
\label{eqAv}
\end{equation}
The $+$ solution was previously presented  in a slightly different form
in  \cite{kessler2014distribution}
and here the question is how to choose the solution for the corresponding
polymer models. 
Clearly, for $d=2$,  $\langle A\rangle^{+} =\langle A\rangle^{-}$, 
 indicating that this is a critical dimension. 
Further 
 $\langle A\rangle^{+}$ in $1$ and $3$   
dimensions are identical and so is $\langle A \rangle^{-}$
as the result of the
symmetry around $d=2$ in Eq. (\ref{eqH}).
The scaling $\langle A\rangle \propto N^{3/2}$ is expected since 
$r$ scales with the square root of $N$ as for Brownian motion, so
the integral over the random processes $r$ scales like $N^{3/2}$. 

We investigate
the physical interpretation of the two possible  boundary conditions.
A mathematical classification  
of
boundary conditions  was provided in 
 \cite{pitman1982decomposition,
martin2011first} and here we find the physical situations where these
conditions apply. We examine the
behavior of the probability current associated with the $k^\mathit{th}$ mode:
$J^\pm_k= -\frac{1}{2}\phi_k^\pm(r_0) \left((r^{(d-1)/2}\phi_k^\pm(r))' + (1-d)r^{(d-1)/2}\phi_k^\pm(r)/r\right)$  for $r$ near the boundary $r_0\to 0$  in
 dimension $d$.
The analysis is summarized in Table \ref{table1}.
We see that in dimension two and higher, the current
on the origin is either zero or positive. A positive current
at the boundary means that probability is flowing into the system,
which  is an unphysical situation in our system. 
Hence we conclude that in dimension two and higher,
the $-$ solution is not relevant.
This implies that statistics of excursion and reflected bridges (and equivalently, ideal and weakly interacting ring polymers) are identical for $d\ge2$ and correspond to the $+$ solution.

\begin{table}[tb]
\centering 
\begin{tabular}{c c c c c} 
 \hline\hline 
$d$&$\phi_k^+$&$J_k^+$&$\phi_k^-$&$J_k^-$\\ [0.5ex] 
\hline 
$d=1$ \quad& \quad$\phi_k^+\sim r$\quad& \quad$J_k^+<0$\quad&
\quad$\phi_k^-\sim \textit{Const}$\quad& \quad$J_k^-=0$\\
$d=2$ \quad& \quad$\phi_k^+\sim \sqrt{r}$\quad& \quad$J_k^+=0$\quad&
\quad$\phi_k^-\sim \sqrt{r}\log r$\quad& \quad$J_k^->0$\\
$d=3$ \quad& \quad$\phi_k^+\sim r$\quad& \quad$J_k^+=0$\quad&
\quad$\phi_k^-\sim \textit{Const}$\quad& \quad$J_k^->0$\\
\hline 
\end{tabular} 
\caption{Behavior of the probability eigenfunctions and probability currents
in the proximity of the origin ($r=0^+$) for the $+/-$ solutions in different
dimensions. A current $J^{-}>0$ on the origin is unphysical hence the critical
dimension where local interactions are uniportant is $2$. }
\label{table1} 
\end{table} 


In Fig. \ref{fig:allD_theory}
 we compare the results of the
simulations of the ideal and weakly interacting polymer models with
our theoretical results for   $P^{+}(A/\langle A \rangle)$, as given in
Eq. (\ref{eqPDF}). 
As noted above, for $d \ge 3$, we see
that even for finite size chains the local interaction is not important, and that the
theory and simulations perfectly match, while for $d=2$ there are strong finite size effects in the 
weakly interacting case. 

{\em Self-avoiding polymers.} 
Extensive simulations of ring SAWs were performed on cubic
lattices.  
As has already been pointed out, the global expansion of a polymer is
characterized by the exponent $\nu_p$.
Since $A$ constitutes a
measure of the overall size of a polymer, its behavior for large $N$ should
follow $A \sim N^{\nu_p + 1}$. 
 For the ideal and weakly-interacting chains $\nu_p=1/2$, and
$A\sim N^{3/2}$.
For SAWs the exact value of the exponent depends on  $d$, i.e.,
$\nu_p=\nu_p(d)$. $\nu_p=1,0.75$, and $0.5$ are  known exactly for
$d=1,2$, and $4$ \cite{nienhuis1982exact}, respectively, while for $d=3$,
$\nu_p$, based on renormalization group
considerations and Monte Carlo simulations, is $\nu_p \simeq0.588$. 
These prediction were extensively
 tested numerically for the observable of interest $A$ with a  critical dimension of $d=4$, characterized by a very slow convergence of the weakly-interacting model (see SM).
        
While the scaling behavior of the SAW model is different from that of the other two models (as reflected in
$\nu _p$), as we have noted, the scaled PDFs, $P(A /\langle A
\rangle)$ are nevertheless similar. 
A  striking  observation is  that the SAWs in $d=2$ and $3$ coincide to the precision of our measurements
with
the comparably narrow PDF of the $d=5$ non-interacting model, 
 (see Fig.  
\ref{fig:allD_theory}).
  That these distribution are 
narrower than the non-self-avoiding case can be qualitatively explained as follows: Since the interaction forbids
many compact conformations, the fluctuations of the area become smaller. This is
easily observed in the extreme case of a linear SAW in one dimension where only
one conformation is allowed and the scaled PDF assumes the form of a
$\delta$-function.

{\em Discussion.} 
The mapping of ring polymer models to the reflected Bessel bridge and 
excursion is very promising since it implies that not only
the observable $A$ can be analytically computed, but also other
measures of statistics of ring polymers. An example would be the maximal 
distance from one of the monomers to any other monomer, 
since that would relate to extreme value
statistics of a correlated process.  The famed Airy distribution
describes both the one dimensional polymer, as well as the three dimensional
one, due to the symmetry we have found in the underlying Hamiltonian.
The case of $d=2$ is critical in the sense that interaction on the origin
becomes negligible, though for finite size chains it is still important.
Boundary conditions of the Feynman-Kac equation were related to physical
models, which allowed as to select the solutions relevant for physical
models.  The SAW polymer exhibits
interesting behavior; the distribution of $A/\langle A \rangle$ is
identical (up to numerical precision) in dimension $2$ and $3$ and
corresponds to the non-interacting models in dimension $5$.
Further  work on this observation is required.

{\em Acknowledgments:}
This work is supported by the Israel Science Foundation (ISF).

\bibliographystyle{rsc}

\newpage
\onecolumngrid
\renewcommand{\thepage}{S\arabic{page}}  
\renewcommand{\thesection}{S\arabic{section}}   
\renewcommand{\thetable}{S\arabic{table}}   
\renewcommand{\thefigure}{S\arabic{figure}}
\renewcommand{\theequation}{S\arabic{equation}}
\setcounter{figure}{0}
\setcounter{equation}{0}
\setcounter{table}{0}

\appendix
\begin{center}
{\Large SUPPLEMENTARY MATERIAL}
\end{center}

\section{Ring Polymer Simulations in $d$ dimensions}
\label{SM_sec1}

\subsection{Ideal and Weakly Interacting Polymers}

In our simulations for the ideal and weakly-interacting ring polymers the
chain is built of $N$ consecutive bonds on a lattice. In the $i$th step, the bond displacement is
 $\Delta r_{i,j}=\pm1$ in each of the $d$ directions, $j=1,\ldots,d$, hence the bond length is  $b=\sqrt{\sum_{j=1}^d(\Delta {r_{j}})^2}=\sqrt{d}$. 
For example, for $d=3$, starting from the origin ($i=0$) with first step of
$\Delta\mathbf{r} = (+1,-1,+1)$ (yielding $b=\sqrt{3}$) we reach the lattice site $\mathbf{r}_1 = (+1,-1,+1)$. For a second step of $\Delta\mathbf{r} = (-1,-1,-1)$ we end up at lattice site $\mathbf{r}_2 =(0,-2,0)$ for the $i=2$ monomer.

In order to maintain the closure condition of the chain, we
choose an array of  length of $N$ with $\ensuremath{N/2}$ components of $(+1)$ and $\ensuremath{N/2}$ of $(-1)$ in each of the 
directions ($d$ such arrays), and then shuffle them for each direction separately. For each $i$, the components of our $d$ dimensional step are the $i$th values of these arrays. The sum of all of the displacements in each direction is then naturally zero
so that the last monomer is always positioned at the origin. For the ideal chain model we built $10^{6}$ such
conformations while for the weakly interacting we threw away all the conformations that crossed the origin prior to the final monomer. For
each of the conformations we calculated $A=\sum_{i=0}^{N}|r_{i}-r_{0}|$,
where $r_{0}=0$, and plotted the distribution of this parameter.

\begin{figure}
\centering
\includegraphics[width=0.55\textwidth]{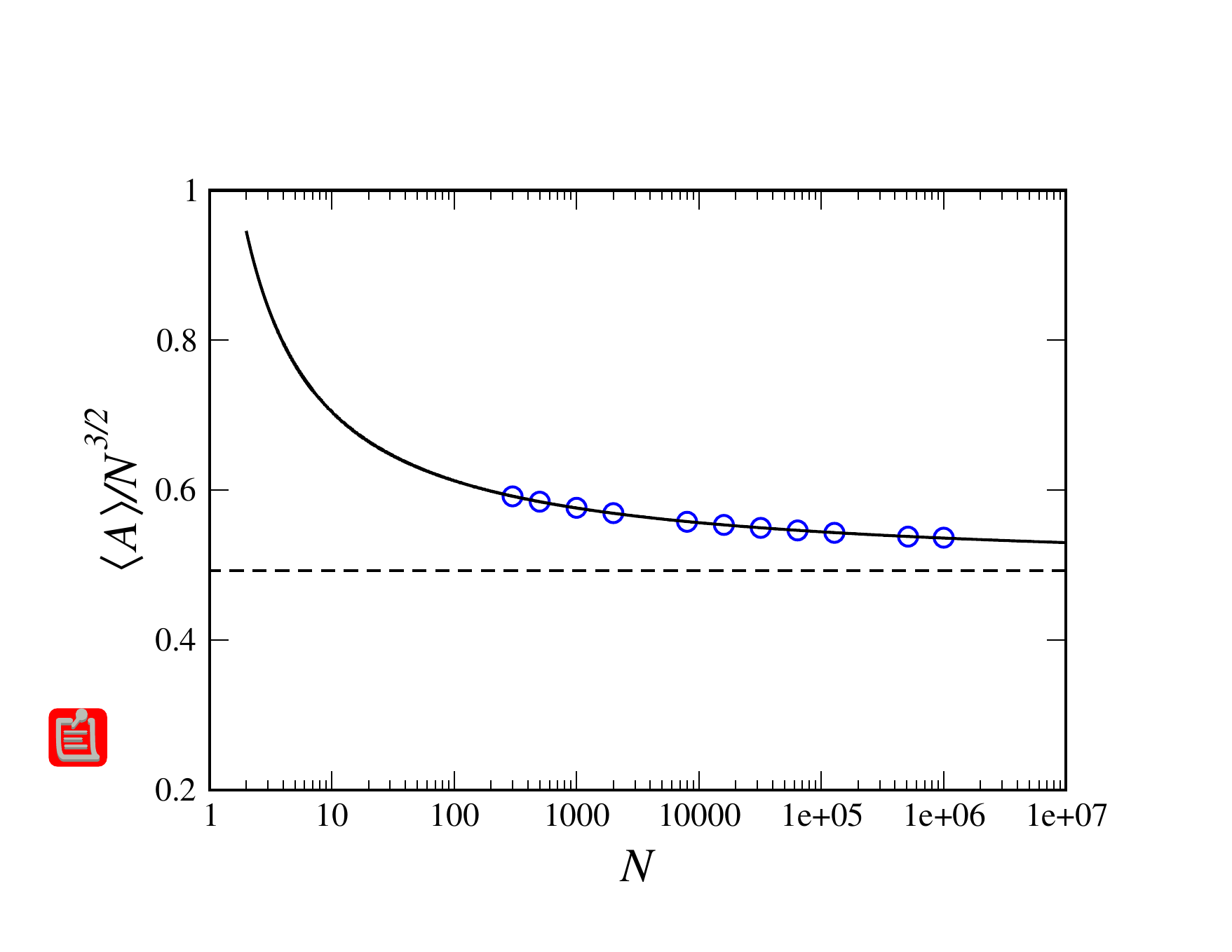}
\caption{(color online) $\langle A\rangle/N^{3/2}$ as a function of $N$ for the critical
dimension $d=2$: weakly-interacting polymer simulations (blue circles) and the fitted
curve of the function: $\langle A\rangle/N^{3/2}\simeq 0.497 + 0.544/\log(N)$. The black
dashed line is the theoretical value of $c^+=0.4922$ for $N\rightarrow\infty$.}
\label{fig7_Convergence}
\end{figure}

According to Eq. (11) in the paper, for $N\rightarrow\infty$ we have
$\langle A \rangle=c_{\pm}N^{3/2}$. By this we can check the convergence
of the simulations to the theory as a function of $N$. At the critical
dimension, $d=2$ this convergence becomes very slow.
In Fig. \ref{fig7_Convergence} we plot $\langle A \rangle/N^{3/2}$ for different
$N$
values (in a logarithmic scale) in $d=2$, where $\langle A
\rangle=c_{+}=0.4922$ is the theoretical value for
$N\rightarrow\infty$. One can observe the very slow logarithmic
convergence to this value.

\subsection{Self-Avoiding Ring Polymers} 
\label{appsub::SAW}

Monte Carlo (MC) simulations have been applied to self-avoiding ring polymer
models \cite{meirovitch1988statistical} on square, simple cubic, and $d=4$
hyper-cubic lattices. The polymer
consists of $N$ monomers (and $N$ bonds), where the first monomer is
attached to the origin of the coordinate system on the lattice, and the $N$th
monomer is the nearest-neighbor to the origin. 
At step $j$ of the MC process, monomer $k$ ($1\leq k\leq N-1$) is selected at
random (i.e., with probability $1/(N-1)$) and the segment of $m$
monomers following $k$ (i.e., $k+1, k+2,…, k+m$) become subject to change
in the MC process; the rest of the chain (i.e., monomers $1$ to $k$ and $k+m+1$
to $N$) is held fixed (notice that if  $k$ is at the end of the chain, $N-m+1 <
k \leq N-1$, $m$ decreases correspondingly from $m-1$ to $1$). Thus, this
current segment is temporarily removed and a scanning procedure is used to
calculate all the possible segment configurations of $m$ monomers 
satisfying the excluded volume interaction and the loop closure condition
(i.e., the segment of $m$ monomers should start at $k$ and its last monomer,
$k+m$ is a nearest neighbor to monomer $k+m+1$; notice that the initial
segment configuration is generated as well). The segment configuration for step
$j$ is chosen at random out of the set of $\mathcal{L}$ configurations generated
by the scanning procedure and the MC process continues. 

This process starts from a given ring configuration, whose transient influence
is
eliminated by a long initial simulation, which leads to typical equilibrium
chain configurations. Then, every certain constant amount of MC steps the
current ring configuration is stored in a file to create a final sample of $n$
rings from which the averages and fluctuations of the physical properties of
interest are calculated.
The segment sizes used are $m=10$ for the square lattice, $m=6$ for the simple
cubic lattice, and $m=4$ for $d=4$. For each lattice several chain lengths, $N$
are studied.

\begin{figure}
\centering
\includegraphics[width=0.55\textwidth]{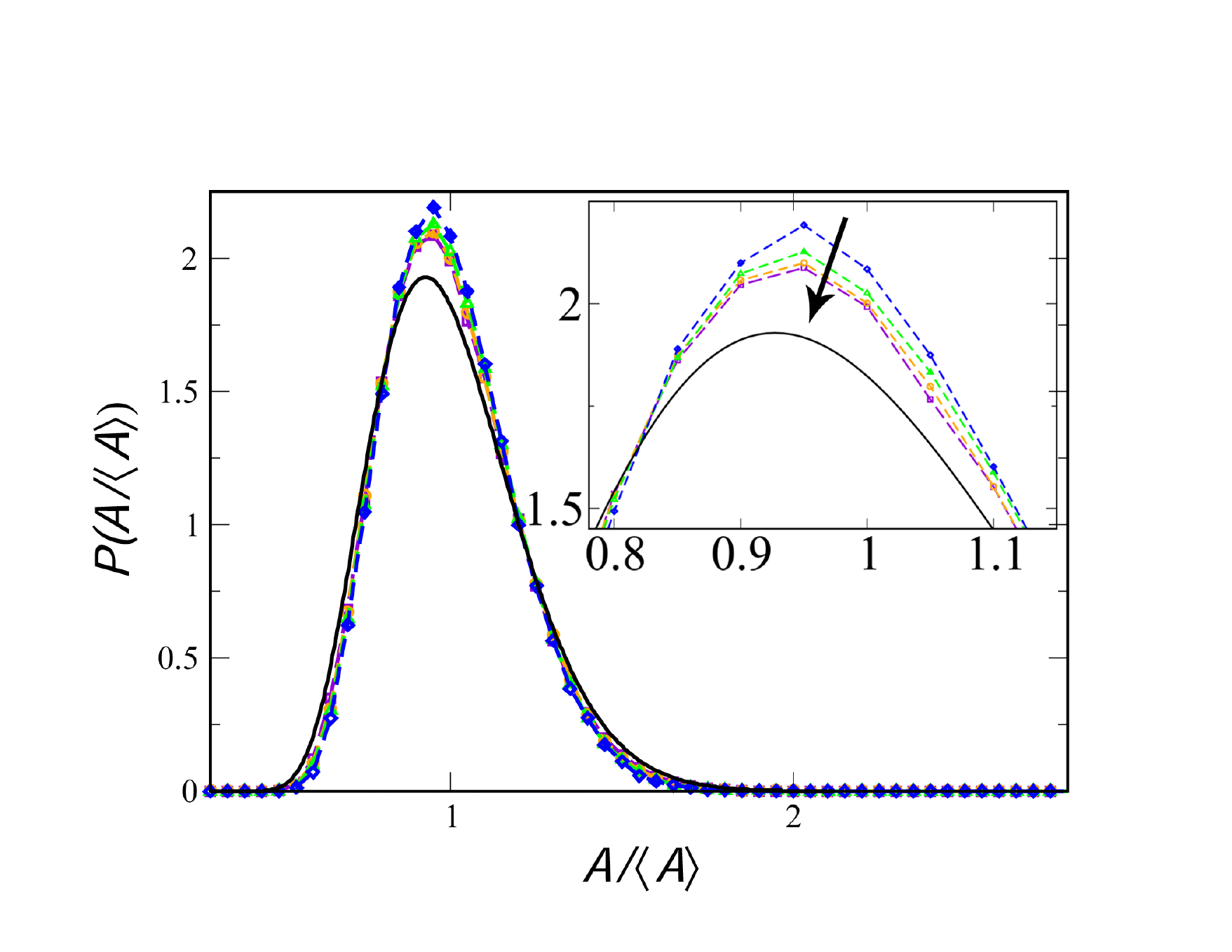}
\caption{(color online) Theoretical $P(A/\langle A\rangle)$ for reflected
Bessel bridges and Bessel excursions for $d=4$ (black solid line) along with
simulations of $d=4$ SAW for $N=200$ (blue diamonds), $N=400$ (green
triangles), $N=800$ (blue circles), $N=1600$ (purple squares).}
\label{fig:d4SAW}
\end{figure}

We calculated the averages of $R_g^2$,
\begin{equation}
R_g^2= \frac{1}{N} \sum_{i=0}^{N-1}{({\bf{r}}_i-{\bf{r}}_{c.m.})^2},
\end{equation}
where
\begin{equation}
{\bf{r}}_{c.m.}\equiv \frac{1}{N} \sum_{i=0}^{N-1}{\bf{r}}_i
\end{equation}
and $A= \sum_{i=0}^{N-1}{|{\bf{r}}_i-{\bf{r}}_0|}$.
For large $N$, $\langle R_g \rangle = \sqrt{\left\langle R_g^2 \right\rangle}$
increases as $\sim N^{\nu_p}$ where $\nu_p$ is a critical exponent and as
discussed in the text, ${\nu_A}={\nu_p+1}$.
However, we are mainly interested in the fluctuations of
$A$, i.e., in the shape of the scaled distribution of $A/\langle A\rangle$ 
for different $N$. To check the reliability of the simulations we provide in
Table \ref{tableAppHagai} the results obtained for $R_g$.
The $\nu_{calculated}$ results for $d=2$ and $d=3$ are equal within the error
bars to those of $\nu_{predicted}$ while for $d=4$ $\nu_{calculated}$ is too
large due to a logarithmic correction to scaling, which would become
insignificant
only for much larger $N$. In fact, considering this correction in the analysis
has led indeed to $\nu_{calculated}\simeq 0.5$. The same quality of results for
${\nu_A}$ has been obtained for the observable $A$ (where $\nu_A\simeq1.5$ for $d=4$). 

\begin{table}[tb]
\centering 
\begin{tabular}{c c c c} 
\hline\hline 
$d$ \qquad&$N$ range\qquad& $\nu_{calculated}$\qquad
& $\nu_{predicted}$\qquad\\ [0.5ex]
\hline 
$2$ \qquad & $200-1600$\qquad & $0.7507\pm0.002$\qquad & $0.75$
(exact)\qquad\\
$3$ & $60-1002$ & $0.5890\pm0.002$ & $0.588$\\
$4$ & $200-2560$ & $0.525\pm0.015$ & $0.5$ (exact)\\
\hline 
\end{tabular} 
\caption{}
\label{tableAppHagai} 
\end{table} 

The $d=4$ SAW case is a critical one, since the critical exponent, $\nu_p$ for
lower dimensions significantly differs from the $\nu_p=1/2$ of
the non-interacting models, and for $d\geq4$ the interactions become
unimportant for $N\rightarrow\infty$. Hence, we expect
$P(A/\langle A\rangle)$ of the $d=4$ SAW to coincide with that of the
non-interacting model. However, for finite $N$ the interaction still has an
effect on the distribution's shape, and an even more pronounced one for ring
polymers.
For the values of $N$ we used in our SAW simulations the curve had not yet
converged as can be seen in Fig. (\ref{fig:d4SAW}). A downward trend  of the
curves towards that of the non-interacting case (i.e. towards convergence) can
nevertheless be seen.  A similar problem is not found for SAW in $d=2,3$ which are reported in the main text.

\section{Numeric values of $\lambda_k$ and $d_k$} 
\label{app::Lambda_kd_k} 

The theoretical PDFs for different dimensions, presented in Eq. (10) in the
paper, may be plotted using  MATHEMATICA$^\circledR$. In order to find the numerical coefficients $\lambda_k$
(eigenvalue) and $d_k$ (the normalization coefficient of the eigenfunction)
values of the $k$th mode, we used the numerical method described in detail in
\cite{barkai2014area}. In Table (\ref{tablel:eigenvalues}) we present the values
of the first few $\lambda_k$ and $d_k$ for different boundaries in different
dimensions. We found that the first $7$ eigenvalues are usually sufficient
for the evaluation of $P^\pm(A/\langle A \rangle)$.

\begin{center}
\begin{table}[tb]
\centering 
\begin{tabular}{c c c c c c c c c  } 
 \hline\hline 
Dimension&Value&$k=1$&$k=2$&$k=3$&$k=4$&$k=5$&$k=6$&$k=7$\\
[0.5ex] %
\hline 
$1,3$&$\lambda_k^+$&$2.3381$&$4.088 $&$5.2056
$&$6.7871$&$7.944$&$9.02265$&$10.040$ \\
 
$\qquad$ &$d_k^+$&$1$&$1$&$1 $&$1$&$1$&$1$&$1$ \\
 \hline\hline
$1$&$\lambda_k^-$&$1.0188$&$3.2482 $&$4.8201
$&$6.1633$&$7.3722$&$8.4885$&$9.5345$  \\

$\qquad$ &$d_k^-$&$0.99088$&$0.5550$&$0.4554
$&$0.4241$&$0.3837$&$0.3663$&$0.3566$ \\ 
\hline 
\hline 
$2$ & $\lambda_{k}^+$ &$ 1.738$ & $3.671$ & $5.170$ &$ 6.475$ & $7.658$ &$
8.755 $&$ 9.787$ \\ 
$\qquad$ & $d_{k}^+$ & $1.1391$ & $0.9195$ &$ 0.8386 $& $0.7885 $& $0.7597$ &$
0.7317$ &$ 0.7109 $\\ 
\hline 
\hline 
$4$ & $\lambda_{k}^+$ & $2.873$ & $4.494$ & $5.868 $& $7.098$ & $8.231$ &$
9.291 $& $10.294$ \\ 
$\qquad$ & $d_{k}^+$ & $0.7585$ & $0.8807$ & $0.9523$ &$ 1.0047$ & $1.0482 $&
$1.0820 $& $1.1108$ \\ 
\hline 
\hline 
$5$ & $\lambda_{k}^+$ & $3.362$ & $4.885$ & $6.208 $& $7.406 $& $8.516 $&
$9.558$ & $10.547 $\\ 
$\qquad$ & $d_{k}^+$ & $0.5187$ & $0.6762$ & $0.7838$ &$ 0.8646$ & $0.9405 $&
$1.0012 $& $1.0531 $ \\ 
\hline 
\end{tabular} 
\caption{The first $7$ eigenvalues, $\lambda_k$ and the corresponding numeric
coefficients $d_k$ required for plotting the theoretical PDFs. The eigenvalues $\lambda_k^+$
of the $+$ solutions for $d=1$ and $d=3$ are the negatives of the zeros of the Airy function: $\textrm{Ai}(-\lambda_k^+)=0$.  
The eigenvalues $\lambda_k^-$ of the $-$ solution are the negatives of the zeros of its derivative: $\textrm{Ai}'(-\lambda_k^-)=0$.  The $d_k^-$ for $d=1$
are tabulated in Ref.~\cite{abramowitz1972handbook}.}
\label{tablel:eigenvalues} 
\end{table}  
\end{center}

%

\end{document}